\documentclass[prl,twocolumn,letterpaper,showpacs,superscriptaddress]{revtex4}

\def\be{\begin{equation}}
\def\fin{\end{equation}}
\def\disp{\displaystyle}

\usepackage[latin1]{inputenc}

\usepackage{graphicx,bm,amssymb,amsmath}
\begin{document}
\title{Bidimensional intermittent search processes: an alternative to  Lévy flights strategies}

\date{\today}

\author{O. Bénichou}
\affiliation{Laboratoire de Physique Th{é}orique de la Mati{è}re Condens{é}e, UMR CNRS 7600,
Universit{é} Pierre et Marie Curie, 4 Place Jussieu, 75252 Paris, France}
\author{C. Loverdo}
\affiliation{Laboratoire de Physique Th{é}orique de la Mati{è}re Condens{é}e, UMR CNRS 7600,
Universit{é} Pierre et Marie Curie, 4 Place Jussieu, 75252 Paris, France}

\author{M. Moreau}
\affiliation{Laboratoire de Physique Th{é}orique de la Mati{è}re Condens{é}e, UMR CNRS 7600,
Universit{é} Pierre et Marie Curie, 4 Place Jussieu, 75252 Paris, France}
\author{R. Voituriez}
\affiliation{Laboratoire de Physique Th{é}orique de la Mati{è}re Condens{é}e, UMR CNRS 7600,
Universit{é} Pierre et Marie Curie, 4 Place Jussieu, 75252 Paris, France}

\begin{abstract}

Lévy flights are  known to be optimal search strategies in the particular case of 
 revisitable targets. In the   relevant situation of 
non revisitable targets,
 we propose  an alternative    model of bidimensional search processes, which  explicitly relies on the widely observed intermittent behavior of foraging animals.   We show analytically that
intermittent strategies can minimize the search time, and therefore do
constitute real optimal strategies.
We study two representative
modes of target detection, and determine which features of the search time are robust
and do not depend on the specific characteristics of detection
mechanisms. In particular, both modes lead to a global minimum of
the search time as a function of the typical times spent in each
state, for the same optimal duration of the ballistic phase. This last quantity  could be a universal feature of bidimensional intermittent search strategies.

\end{abstract}


\maketitle


Search processes, involving a searcher and a target of unknown position,
 play 
an important role in many physical, chemical or biological problems. This is for instance the case of reactants  diffusing in a solvent until they get close enough  to react \cite{rice}, or of a  protein  searching for its  specific target site on DNA \cite{berg81,halford,nousproteine}. One can also mention animals searching for  food\cite{bell,obrien,visw,viswprl,viswprl2,nousanimaux}  
or coast-guards trying to locate wreck victims\cite{coast}. In all these examples, 
it is of great  importance to minimize the search time. Since the pioneering works
of Viswanathan et al \cite{visw}, the question of determining
optimal search strategies has appealed a  growing attention
\cite{viswprl,viswprl2,metzler,szabo,sokolov,slutsky,klafter1,klafter,levitz}.

In this context, Lévy flights strategies  have been proved  to  play a
crucial role in  such optimization problems\cite{visw,viswprl,viswprl2}. However, two limitations of these strategies have to be mentioned. First, 
 Lévy flights trajectories have been shown to
optimize the search efficiency, but only in the  particular case where the
targets are regenerated at the same location after a finite time
\cite{visw,viswprl2}, which can not be taken as a general rule. Indeed, in the
 case of destructive search where each target can
be found only once, or in the case of a single target, the optimal
strategy proposed in \cite{visw} is not anymore of Lévy type, but reduces to a linear ballistic
motion.  Second, 
as for  the applications to behavioral ecology,  the destructive search is relevant to many situations\cite{bell,obrien}. However, the  purely ballistic strategy predicted by \cite{visw} in that case 
can  not account 
for the generally observed reoriented animal trajectories\cite{bell}.

Alternatively to these Lévy strategies, it has been observed that 
intermittent search strategies are widely used by foraging animals \cite{obrien,kramer}. Many searchers
combine phases of fast displacement, non receptive to the
targets, and slow reactive search phases.  Everyday-life examples also
confirm that we instinctively adopt such intermittent behavior when looking for 
a lost object: we search carefully around one location, then move quickly
to another unvisited area and then search again.

Up to now, only 1D
models of such intermittent search have been developed, providing a
satisfactory agreement with experimental data from behavioral
ecology \cite{nousanimaux}. Here we develop a model of 2D
intermittent search strategies,  which encompasses a  much broader field
of applications, in particular for animal or human searchers. We
show that bidimensional intermittent search strategies
 {\it do optimize} the search time for  non revisitable targets. We explicitly determine how to share the time between the phases of non reactive displacement and of reactive search to
find a target in the quickest way.
 From a technical point of view, we also obtain as a by-product the mean first passage time
for a Pearson type random walk, which  belongs to a  class 
of non trivial problems  which  have been investigated
for long \cite{Pearson,Blanco,Mazzolo,Nousepl,Satya}. 
Our approach relies on  an approximate
analytical solution based on a decoupling hypothesis, which proves
to reproduce quantitatively our numerical simulations over a wide
range of parameters.

\begin{figure}
\scalebox{0.18}{
\includegraphics{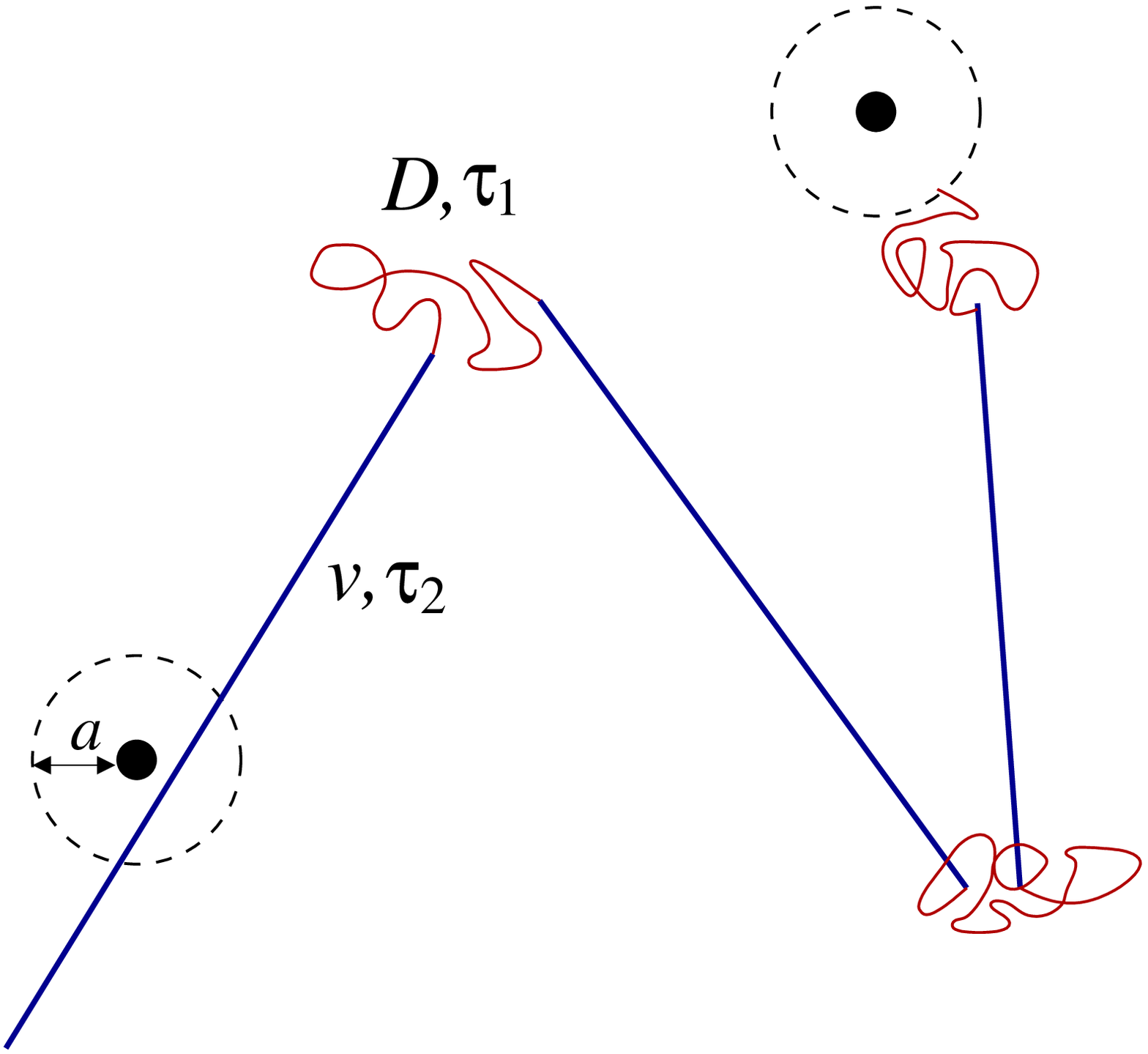}}
\hspace {1.5cm}
\scalebox{0.18}{
\includegraphics{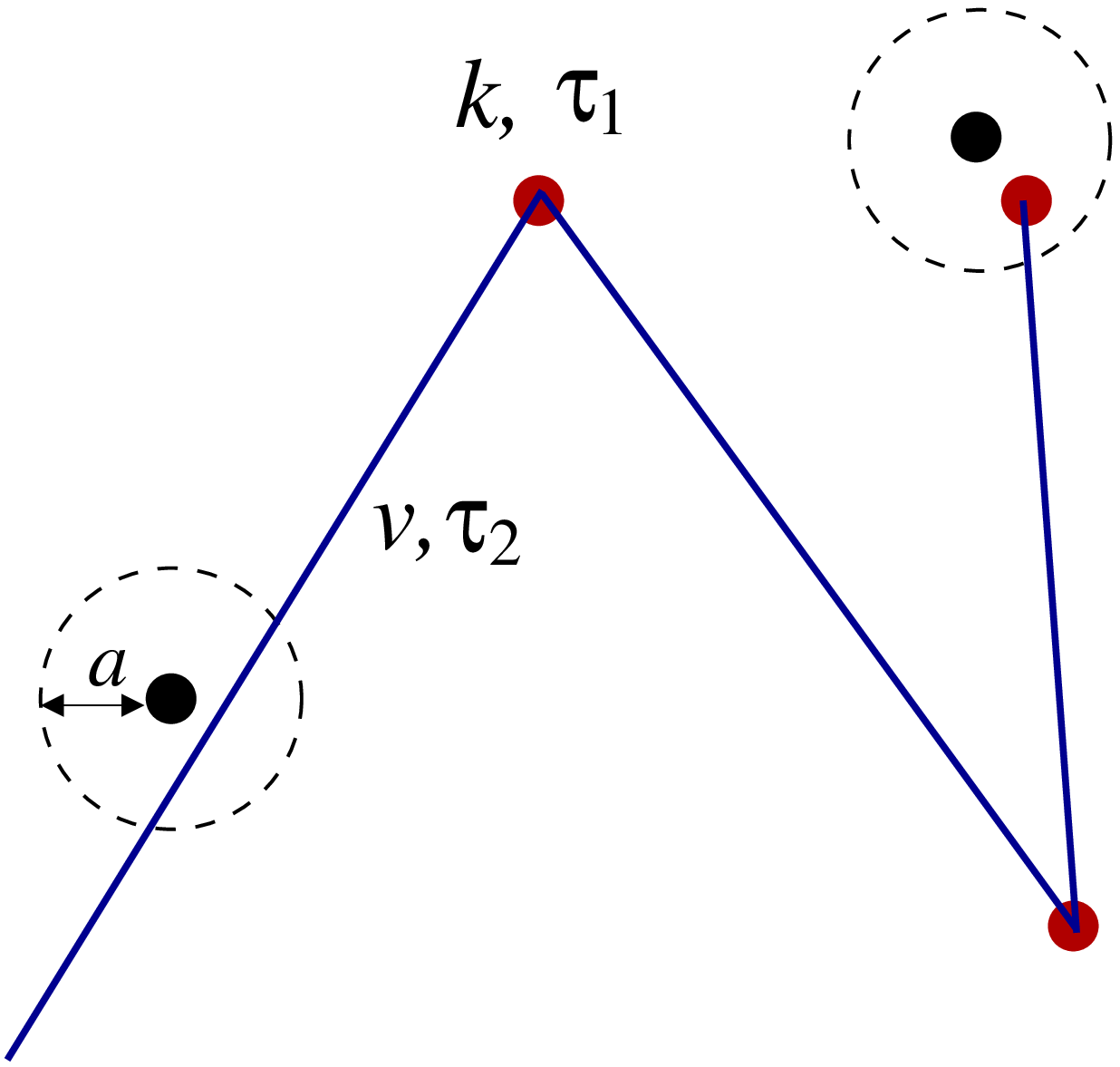}}
\caption{Two models of intermittent search: The searcher alternates slow reactive phases (regime 1)
of mean duration $\tau_1$, and fast non reactive ballistic phases (regime 2) of mean duration $\tau_2$.
 {\it Left}: the slow reactive phase is diffusive and detection is infinitely efficient. {\it Right}: the slow reactive phase
 is static and detection takes place with
finite rate $k$.}
\label{figmodel}
\end{figure}

Following \cite{nousanimaux}, we consider a two state searcher (see Fig.\ref{figmodel}) of
position $\bf r$ that performs slow reactive phases (denoted 1),
 randomly interrupted by fast relocating ballistic flights of
constant velocity $v$ and random direction (phases 2). We assume the
duration of each phase $i$ to be exponentially distributed with mean
$\tau_i$. As fast motion usually strongly degrades perception
abilities\cite{obrien,kramer}, we consider that the searcher is able to find a target
only during  reactive  phases 1.  The detection phase involves
complex biological processes that we do not aim at modeling accurately here.
However, essentially two modes of detection can be put forward, and
lead to  distinguish between two types of reactive phases 1. The
first one, referred to in the following as the "dynamic mode",
corresponds to a diffusive modeling (with diffusion coefficient $D$)
of the search phase as recently proposed in \cite{nousanimaux} in
agreement with observations for vision\cite{huey}, tactile sense or
olfaction\cite{bell}.  The detection is assumed to be  infinitely
efficient in this mode : a target is found as soon as the
searcher-target distance is smaller than the reaction radius $a$. On
the contrary, in the second mode, denoted as the "static mode", the
reaction takes place with a finite rate $k$, but the searcher is
immobile during search phases. Note that this description is
commonly adopted in reaction-diffusion systems\cite{rice} or operational
research\cite{coast}. A more realistic description is obtained by combining both
modes and considering a diffusive searcher with diffusion
coefficient $D$ and finite reaction rate $k$. In order to reduce the
number of parameters and to extract the main features of each mode,
we study them separately by taking successively the limits
$k\to\infty$ and $D\to0$ of this general case. More precisely, in
these two limiting cases, we address  the following questions  :
what is the mean time it takes the searcher to find a target?   Can
this search time be minimized ? And if so for  which values of the
average durations $\tau_i$ of each phase?

We now present the basic equations combining the two  search modes
introduced above  in the case of a  point-like target centered in a spherical
domain of radius $b$ with reflecting boundary.  Note that this
geometry mimics both relevant situations of a single target and of
infinitely many regularly spaced non revisitable targets.
For this process, the mean first passage time (MFPT) at a target
satisfies the following backward equation\cite{redner}:
\begin{equation}\label{back1}
D\nabla^2_{\bf r}t_1+\frac{1}{2\pi\tau_1}\int_{0}^{2\pi}(t_2-t_1)d\theta_{\bf v}-k{\rm I}_a({\bf r})t_1=-1
\end{equation}
\begin{equation}\label{back2}
{\bf v}\cdot\nabla_{\bf r}t_2-\frac{1}{\tau_2}(t_2-t_1)=-1
\end{equation}
where $t_1$ stands for the MFPT starting from state 1 at position ${\bf r}$, and $t_2$ for the MFPT starting from state 2 at
position ${\bf r}$ with velocity ${\bf v}$. ${\rm I}_a({\bf r})=1$ if $|{\bf r}|\le a$ and  ${\rm I}_a({\bf r})=0$ if $|{\bf r}|> a$.
In the present form, these integro-differential equations do not seem to allow for an exact resolution with standard methods.
We propose here an approximate resolution based on the introduction of the following auxiliary functions :
\begin{equation}
s({\bf r})=\frac{1}{2\pi}\int_{0}^{2\pi}\!t_2d\theta_{\bf v},\; {\bf d}({\bf r})=\frac{1}{2\pi}\int_{0}^{2\pi}\!t_2{\bf v}d\theta_{\bf v}.
\end{equation}
Averaging  Eq.(\ref{back2}) and Eq.(\ref{back2}) times ${\bf v}$ over $\theta_{\bf v}$,   one successively gets
\begin{equation}\label{eqs}
{\bf \nabla}\cdot{\bf d}-\frac{1}{\tau_2}(s({\bf r})-t_1)=-1\;{\bf d}=\frac{\tau_2}{2\pi}\int_{0}^{2\pi}({\bf v}\cdot\nabla t_2){\bf v} d\theta_{\bf v},
\end{equation}
which gives in turn
\begin{equation}
{\bf \nabla}\cdot{\bf d}=\frac{\tau_2}{2\pi}\sum_{i,j}\frac{\partial^2}{\partial x_i\partial x_j}\langle v_i v_j t_2\rangle_{\theta_{\bf v}}
\end{equation}
where $ \langle\cdot\rangle_{\theta_{\bf v}}$ stands for the average over $\theta_{\bf v}$.
We now make the following decoupling assumption
\begin{equation}\label{dec}
\langle v_i v_j t_2\rangle_{\theta_{\bf v}}\simeq\langle v_i v_j\rangle_{\theta_{\bf v}}\langle t_2\rangle_{\theta_{\bf v}}=\frac{v^2}{2}\delta_{ij}s({\bf r})
\end{equation}
which leads, together with Eq.(\ref{eqs}), to the diffusion-like equation:
\begin{equation}\label{dtilde}
{\widetilde D}\nabla^2 s({\bf r})-\frac{1}{\tau_2}(s({\bf r})-t_1)=-1
\end{equation}
where ${\widetilde D}=v^2\tau_2/2$.
Rewriting Eq.(\ref{back1}) as
\begin{equation}\label{Dt1}
D\nabla^2 t_1+\frac{1}{\tau_1}(s({\bf r})-t_1)-k{\rm I}_a({\bf r})t_1=-1,
\end{equation}
Eqs.(\ref{dtilde}) and (\ref{Dt1}) together with vanishing normal derivatives at $|{\bf r}|=b$ provide a closed system for the
variables $s$ and $t_1$, whose resolution is lengthy but standard. The validity domain
of assumption (\ref{dec}) is much broader than the  "Brownian" limit
$v\to \infty$ and $\tau_2\to 0$ with ${\tilde D}$ fixed, in which  $t_2$ is
 independent of the direction of ${\bf v}$. Indeed, it is also valid in the limit $v\tau_2\gg b$, in which
 a ballistic phase includes many  reorientations due to successive reflections on the boundary $r=b$. In addition,  
 it can be shown that in one dimension this assumption is
exact.

\begin{figure}
\scalebox{0.17}{
\includegraphics{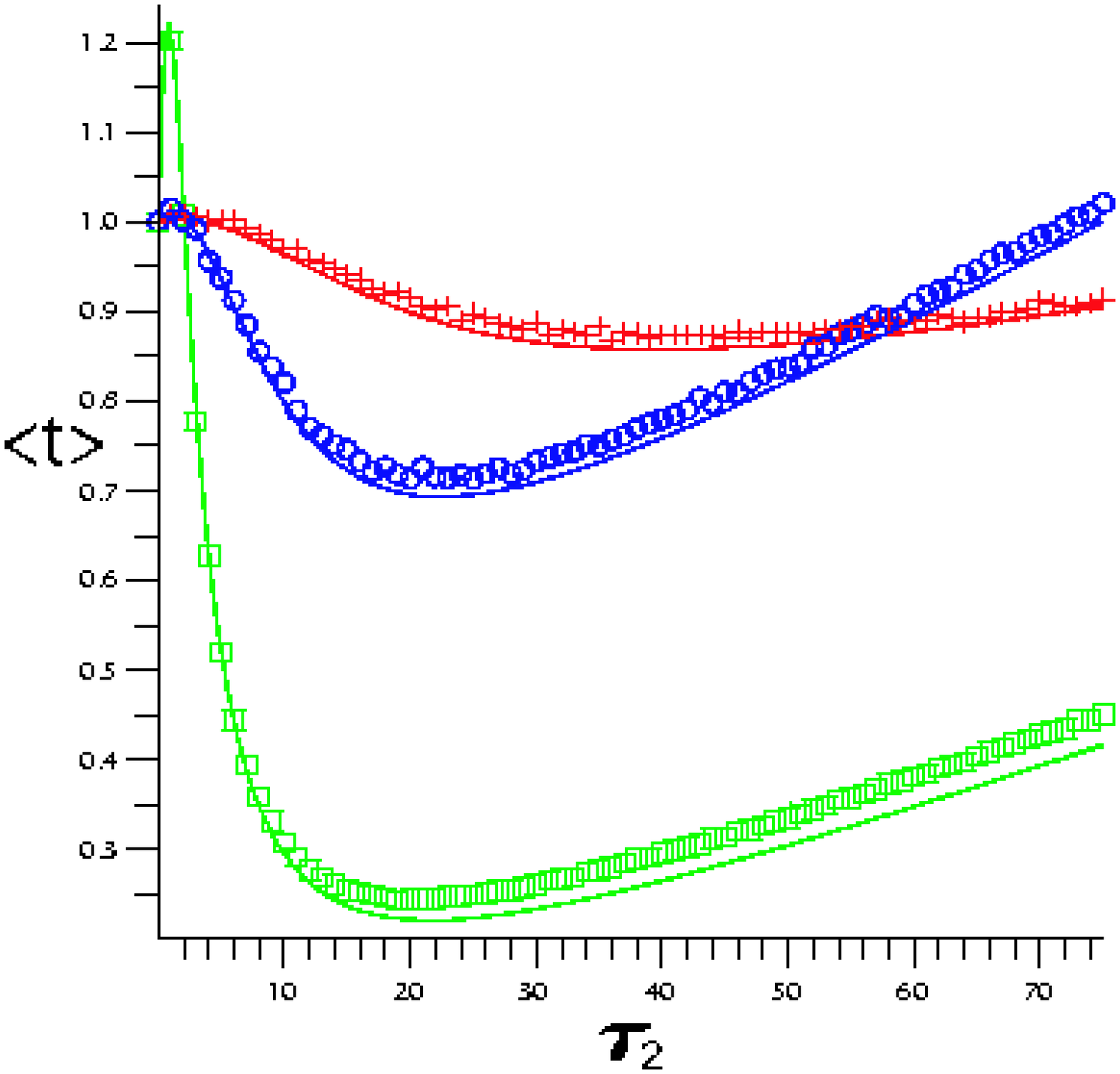}
\hspace{2cm}\includegraphics{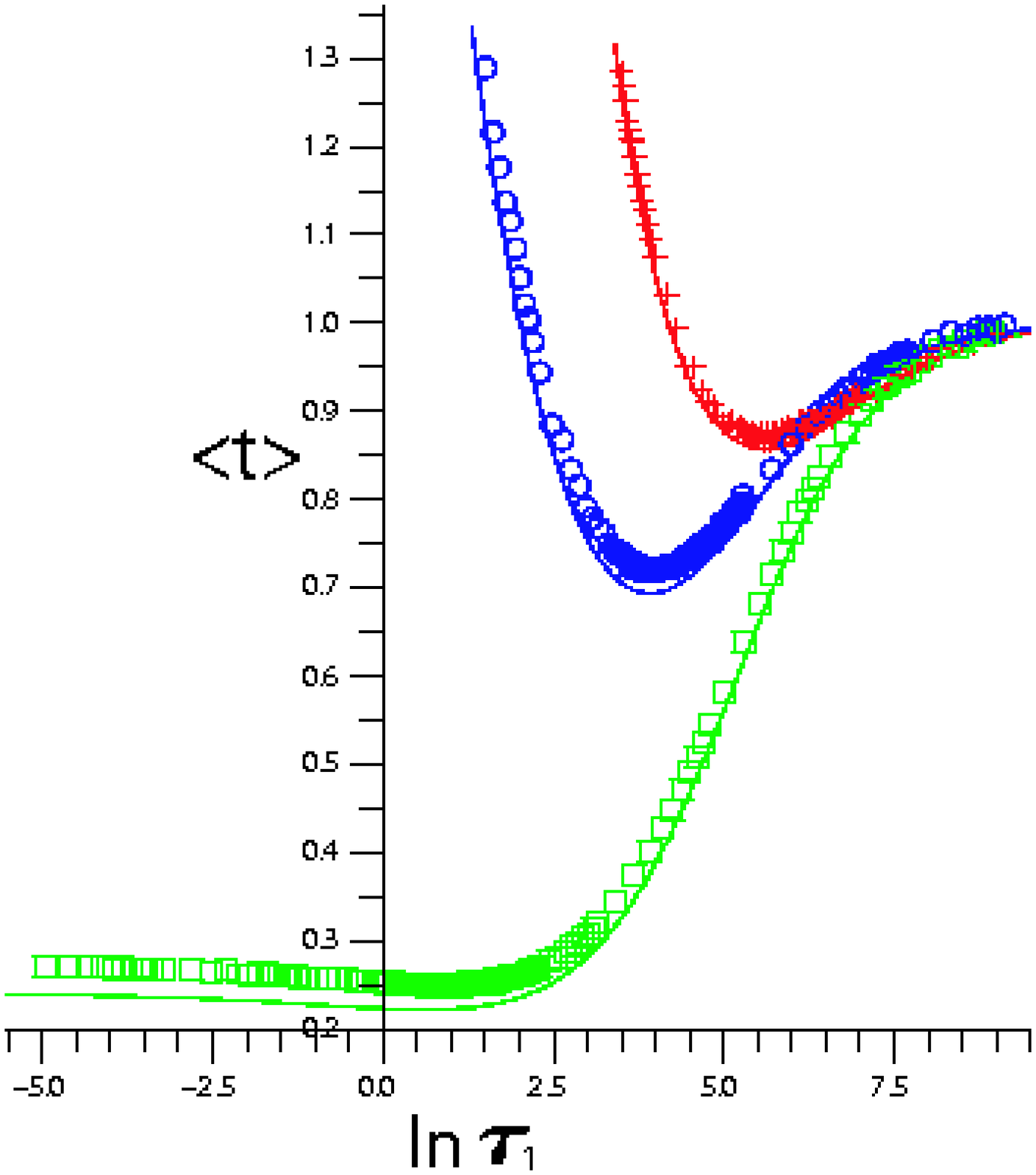}}
\caption{Simulations (points) versus analytical approximate (line)    of the
search time in the "dynamic mode": the search time rescaled by the value in absence of intermittence  as a function of $\tau_2$ (left) and $\ln\tau_1$ (right) (the logarithmic scale has been used due to the flatness of the minimum),  for $D=1$, $v=1$, $b=451$.  {\it Left}:  $a=10$, $\tau_1=1.88$ (green boxes); $a=1$,     $\tau_1=50$ (blue circles); $a=0.1$,    $\tau_1=250$ (red crosses). {\it Right}:  $a=10$, $\tau_2=21.6$ (green boxes); $a=1$,     $\tau_2=22.5$ (blue circles); $a=0.1$,    $\tau_2=41.6$ (red crosses). }
\label{fig2}
\end{figure}

We first present the solution  of Eqs(\ref{dtilde},\ref{Dt1})  in
the "dynamic mode" ($k\to\infty$). The search time $\langle
t\rangle$, defined as $t_1$ uniformly  averaged over the initial
position of the searcher (note that this last averaging reflects the
complete ignorance of the target position), reads in this case:
\begin{widetext}
\begin{equation}\disp\label{tmap}
\langle t\rangle =(\tau_1+\tau_2)\frac{\disp1-a^2/b^2}{\disp(\alpha^2 D\tau_1)^2}\left\{\disp a\alpha(b^2/a^2-1)\frac{\disp M}{\disp 2L_+}-\frac{\disp L_-}{\disp L_+}-\frac{\disp \alpha^2D\tau_1}{\disp 8{\widetilde D}\tau_2}\frac{\disp(3-4\ln(b/a))b^4-4a^2b^2+a^4}{\disp b^2-a^2}\right\}
\end{equation}
\begin{equation}
{\rm with}\  L_\pm={\rm I}_0(a/\sqrt{{\widetilde D}\tau_2})\left({\rm I}_1(b\alpha){\rm K}_1(a\alpha)- {\rm I}_1(a\alpha){\rm K}_1(b\alpha)  \right)\pm \alpha\sqrt{{\widetilde D}\tau_2}\;{\rm I}_1(a/\sqrt{{\widetilde D}\tau_2})\left({\rm I}_1(b\alpha){\rm K}_0(a\alpha)+ {\rm I}_0(a\alpha){\rm K}_1(b\alpha)  \right)\nonumber
\end{equation}

\begin{equation}
{\rm and }\  M={\rm I}_0(a/\sqrt{{\widetilde D}\tau_2})\left({\rm I}_1(b\alpha){\rm K}_0(a\alpha)+ {\rm I}_0(a\alpha){\rm K}_1(b\alpha)  \right)-4\frac{a^2\sqrt{{\widetilde D}\tau_2}}{\alpha(b^2-a^2)^2}{\rm I}_1(a/\sqrt{{\widetilde D}\tau_2})\left({\rm I}_1(b\alpha){\rm K}_1(a\alpha)- {\rm I}_1(a\alpha){\rm K}_1(b\alpha)  \right)\nonumber
\end{equation}
\end{widetext}
where $\alpha=(1/(D\tau_1)+1/({\widetilde D}\tau_2))^{1/2}$, and ${\rm I}_i$ and ${\rm K}_i$  are modified Bessel functions. This
expression (\ref{tmap}) has proved to be in very good  agreement
with numerical simulations for a wide range of  the parameters (see
Fig.(\ref{fig2})).
The optimization of the explicit expression  (\ref{tmap}) leads to
simple forms in the  following situations, depending on the relative
magnitude of the three characteristic lengths of the problem
$a,b,D/v$. We limit ourselves to the the case of low target density
($a\ll b$), which is the  most relevant for hidden targets search
problems. Three regimes then arise.
In the first regime $a\ll b\ll D/v$, the relocating phases are not efficient and intermittence is useless.
 In the second regime $a\ll D/v \ll b$,  it can be shown 
that the intermittence can significantly speed up the search
(typically by a factor 2), but that it does not change the order of magnitude
of the search time.  On the contrary, in the last regime $D/v\ll a\ll
b$, the optimal strategy, obtained for
\begin{equation}\label{grandv}
\tau_{1,\rm min}\sim \frac{D}{2v^2}\frac{\ln^2 (b/a)}{2\ln (b/a) -1}, \;\tau_{2, \rm min}\sim \frac{a}{v}(\ln(b/a)-1/2)^{1/2},
\end{equation}
leads to a qualitative change of the search time which can be rendered arbitrarily smaller than the non intermittent
 search time when $v\to\infty$. This optimal strategy  corresponds to a scaling law
\begin{equation}
\frac{\tau_{1, \rm min}}{\tau_{2, \rm min}^2}\sim \frac{D}{a^2}\frac{1}{\left(2-1/\ln(b/a)\right)^2}
\end{equation}
which here does not depend on $v$. 

\begin{figure}
\scalebox{0.15}{
\includegraphics{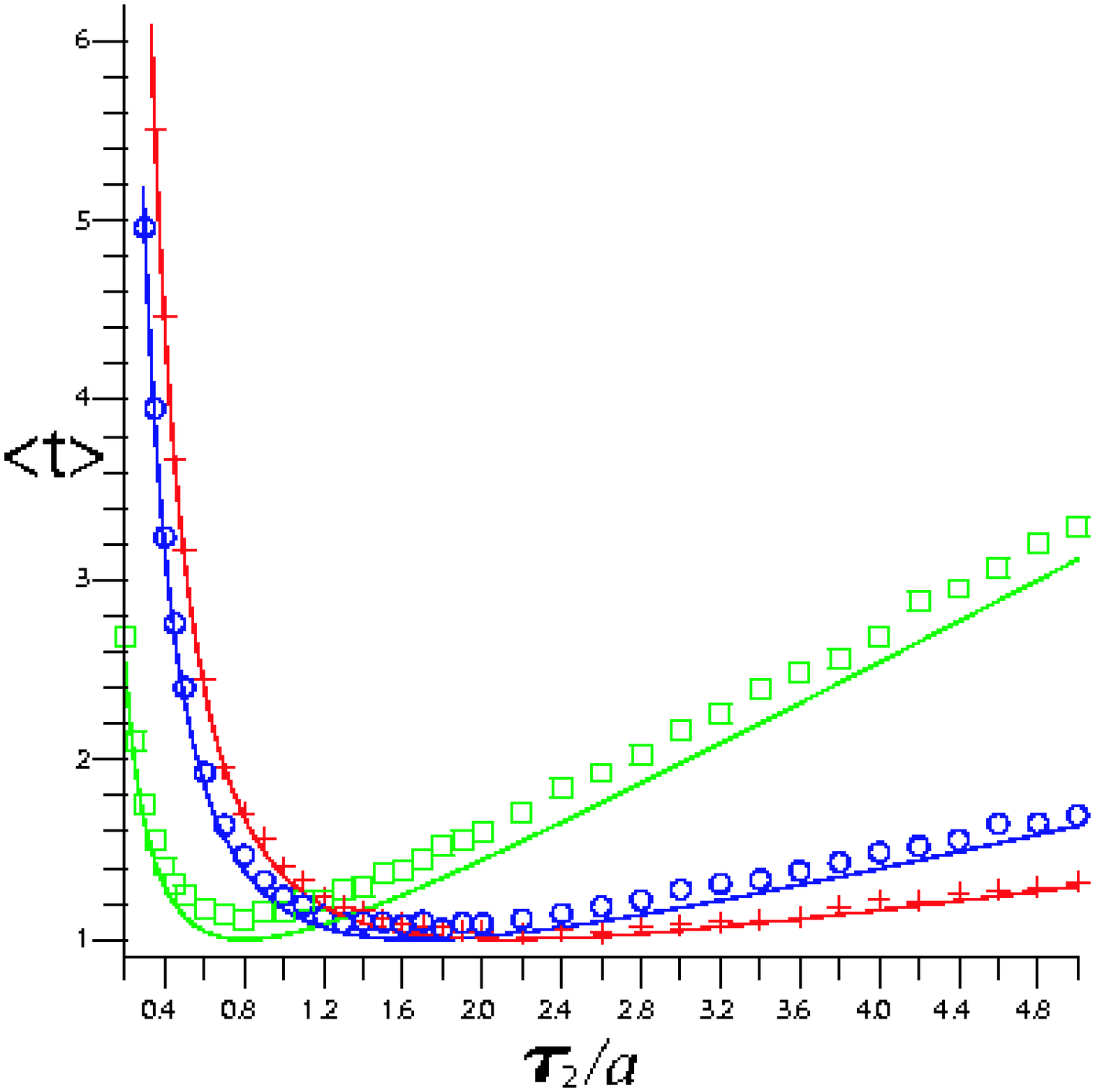}
\hspace {2cm}
\includegraphics{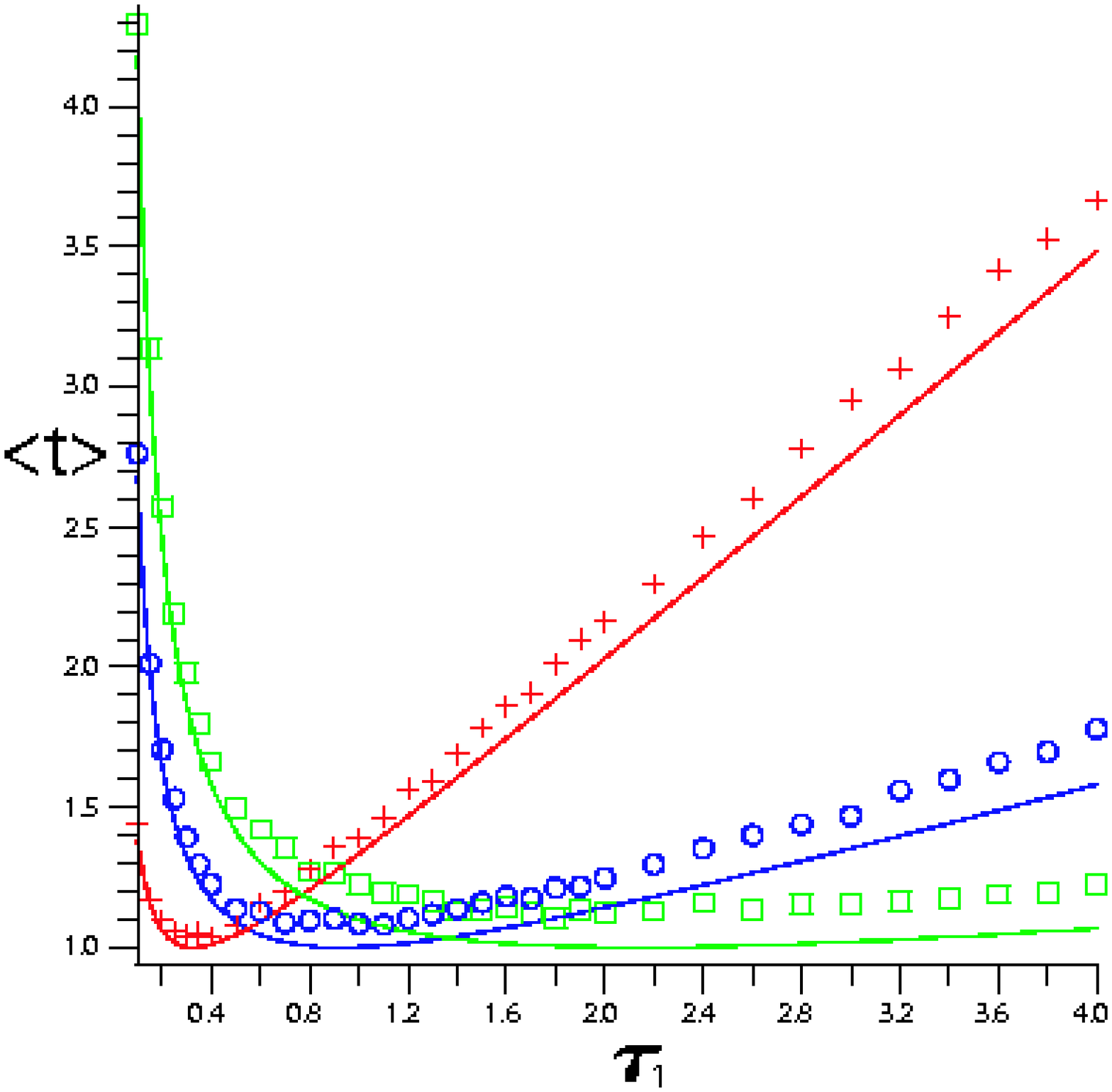}}
\caption{Simulations (points) versus analytical approximate (line)    of the
search time in the "static mode": the search time rescaled by the optimal value  as a function of $\tau_2/a$ (left) and $\tau_1$ (right),  for $k=1$, $v=1$, $b=28$. {\it Left}:  $a=10$, $\tau_1=2.14$ (green boxes); $a=1$,     $\tau_1=0.9$ (blue circles); $a=0.1$,    $\tau_1=0.3$ (red crosses). {\it Right}:  $a=10$, $\tau_2=8$ (green boxes); $a=1$,     $\tau_2=1.7$ (blue circles); $a=0.1$,    $\tau_2=0.22$ (red crosses). }
\label{fig3}
\end{figure}

We now turn to the "static mode" ($D\to 0$), which leads to the following expression for the search time
\begin{widetext}
\begin{equation}\label{searchtime2}
\langle t\rangle = \frac{\tau_1+\tau_2}{2k\tau_1 y^2}\left\{\frac{1}{x}(1+k\tau_1)(y^2-x^2)^2\frac{{\rm I}_0(x)}{{{\rm I}_1(x)}}
+\frac{1}{4}\left[8y^2+(1+k\tau_1)\left(4y^4\ln(y/x)+(y^2-x^2)(x^2-3y^2+8)\right)\right]\right\}
\end{equation}
\end{widetext}
\begin{equation}
{\rm where}\;x=\sqrt{\frac{2k\tau_1}{1+k\tau_1}}\frac{a}{v\tau_2} \;{\rm and}\;y=\sqrt{\frac{2k\tau_1}{1+k\tau_1}}\frac{b}{v\tau_2}
\end{equation}
Here again, this
expression (\ref{searchtime2}) is in very good  agreement
with numerical simulations for a wide range of  the parameters (see
Fig.(\ref{fig3})).
In that case, intermittence is trivially necessary to find the target, and the optimization of the search time (\ref{searchtime2})
 leads for $b\gg a$ to
\begin{equation} \label{statique}
\tau_{1,{\rm min}}=\left(\frac{a}{vk}\right)^{1/2}\left(\frac{2\ln(b/a)-1}{8}\right)^{1/4},\;
\end{equation}
\begin{equation}\label{statique2}
\tau_{2,{\rm min}}=\frac{a}{v}\left(\ln(b/a)-1/2\right)^{1/2},
\end{equation}
which corresponds to the scaling law $\tau_{2,{\rm min}}=2k\tau_{1,{\rm min}}^2$, which still does not 
depend on $v$.

The main results (\ref{grandv}) and
(\ref{statique}), (\ref{statique2})  obtained in the two  modes of search lead to extract the following  remarkable characteristics of intermittent search processes:
(i) In both
cases the search time $\langle t\rangle$ presents a global minimum
for finite values of the $\tau_i$, which means that intermittence is
an optimal strategy. (ii) A very striking and non intuitive feature is that both modes of
search lead to the {\it same optimal value} of $\tau_{2, {\rm min}}$. As this optimal time  does not depend on
the specific characteristics $D$ and $k$ of the search mode, it seems to constitute a general property of intermittent search strategies. (iii)
The optimal  $\tau_{1, {\rm min}}$ are different and depend
explicitly on $D$ and $k$, leading to different scaling laws which
are susceptible to discriminate between the two search modes .

Finally we remark that this model provides as a by-product an approximation for the MFPT for a Pearson type  random walk in the spherical geometry previously defined:
the searcher performs ballistic flights reoriented at exponentially distributed  times,  and, as opposed to standard Pearson walks,  the target can be found only
when the distance between the target and a reorientation point is less than $a$.  This quantity, obtained here 
straightforwardly by taking $k\to\infty$ and $\tau_1\to 0$  in Eq.(\ref{searchtime2}), writes:
\begin{eqnarray}\label{Pearson}
\langle t \rangle =\frac{v\tau_2^2}{4b^2}\left(\sqrt{\frac{2}{a}}(b^2-a^2)\frac{{\rm I}_0(a\sqrt{2}/v\tau_2)}{{\rm I}_1(a\sqrt{2}/v\tau_2)}+\right.\nonumber\\
+\left.\frac{1}{v^3\tau_2^3}(b^4\ln(b/a)+(b^2-a^2)(a^2-3b^2+4v^2\tau_2^2)\right)
\end{eqnarray}
 To our knowledge,  a similar result for standard Pearson walks is still missing. Note that in the limit $v\to\infty$, $\tau_2\to0$ with ${\widetilde D}=v^2\tau_2/2$ fixed, the 
 approximate expression (\ref{Pearson}) gives back the well known  exact expression for the MFPT of a Brownian particle between concentric spheres\cite{redner}. Moreover, for $b\gg a$, 
 the search time (\ref{Pearson}) is minimized again for the same value   (\ref{grandv}) and
(\ref{statique2}) of $\tau_2$, in agreement  with the limit $k\to\infty$ of Eqs. (\ref{statique}),  
(\ref{statique2}).

To conclude, we have proposed a two state model of search processes
for non  revisitable targets, which closely relies on the
experimentally observed intermittent strategies adopted by foraging animals. Using a decoupling
approximation numerically validated,   we have shown analytically
that in the physically most relevant bidimensional geometry,
intermittent strategies minimize the search time, and therefore
constitute  optimal strategies, as opposed to Lévy flights which are
optimal only for revisitable targets. 
We studied two representative modes of search, and determined which
features of the corresponding optimal strategies are robust and do not depend on the specific
characteristics of the search mode. In particular both modes lead to
a global minimum of the search time as a function of the typical
times spent in each state, and the optimal duration of the ballistic relocation
phase is the same for these both modes. As this last time does not depend on the nature of the search mode, 
it could be a universal feature of bidimensional intermittent search strategies.

\end{document}